\documentstyle[12pt]{article}
\renewcommand{\theequation}{\arabic{section}.\arabic{equation}}
\begin{document}
%
%
%
%
\pagestyle{empty}
\vspace* {18mm}
\renewcommand{\thefootnote}{\fnsymbol{footnote}}
\begin{center}
   {\bf FOUR STATE MODELS AND CLIFFORD ALGEBRAS}
   \\[25MM]
   Haye Hinrichsen
   \footnote{e-mail: \tt hinrichs@aster.physik.fu-berlin.de} \\[10mm]
   {\it Freie Universit\"{a}t Berlin\\
   Institut f\"ur Theoretische Physik\\ Arnimallee 14,
   D-14195 Berlin, Germany}
   \\[4cm]
{\bf Abstract}
\end{center}
\renewcommand{\thefootnote}{\arabic{footnote}}
\addtocounter{footnote}{-1}
\vspace*{2mm}
%
With appropriate boundary conditions the anisotropic $XY$ chain in
a magnetic field is known to be invariant under quantum group
transformations. We generalize this model defining a class
of integrable chains with
several fermionic degrees of freedom per site.
In order to maintain the quantum group symmetry
a general condition on the parameters of these systems is derived.
It is shown that the
corresponding quantum algebra is a multi-parameter deformation
of the Clifford algebra. Discussing a special physical example
we observe a new type of zero modes.
\vspace{3.5cm}
\begin{flushleft}
   cond-mat/9404034 \\
   April 1994 \\
\end{flushleft}
\thispagestyle{empty}
\mbox{}
%
%
\newpage
\setcounter{page}{1}
\pagestyle{plain}
\section{Introduction}
\setcounter{equation}{0}
\def\id{{\bf 1}}

In statistical mechanics the anisotropic
$XY$ chain is one of the simplest exactly solvable
models. Its $L$-site
Hamiltonian with periodic boundary conditions
\begin{equation}
\label{PeriodicHamiltonian}
H^{XY}_{per}(\eta,h) \; = \;
-\frac{1}{2}\sum_{j=1}^{L}\:
\left( \,\eta\:\sigma_{j}^{x}\sigma_{j+1}^{x}
\; + \; \eta^{-1}\sigma_{j}^{y}\sigma_{j+1}^{y}
\right)
\;-\; h\, \sum_{j=1}^{L}\; \sigma_j^z
\end{equation}
depends on two parameters, namely
the anisotropy parameter $\eta$ and the
magnetic field $h$ ($\sigma^{x,y,z}_j$
are Pauli matrices acting on site $j$).
This Hamiltonian has a long history
\cite{Nijs,McCoy} and provides
a good model for Helium adsorbed on metallic
surfaces. It also describes the master equation of the kinetic
Ising model \cite{Siggia} and
plays a role in one-dimensional reaction-diffusion
processes \cite{Chemie}.
\\
\indent
The present work is based on the investigation of the anisotropic
$XY$ chain with a {\it special kind of boundary conditions}
defined by the Hamiltonian
\begin{equation}
\label{TwoParameterHamiltonian}
\label{xy}
H^{XY}(\eta,q) \; = \;
-\frac{1}{2}\sum_{j=1}^{L-1}\:
\left( \,\eta\:\sigma_{j}^{x}\sigma_{j+1}^{x}
\; + \; \eta^{-1}\sigma_{j}^{y}\sigma_{j+1}^{y}
\; + \; q\,\sigma_{j}^{z} \; + \;
q^{-1}\sigma_{j+1}^{z} \right) \;,
\end{equation}
\noindent
where $q$ is related to the magnetization by $2h=q+q^{-1}$
(notice that compared to Eq. (\ref{PeriodicHamiltonian}) there are
additional surface fields at the ends of the chain).
These boundary conditions make the system to be invariant under
quantum group transformations \cite{Saleur,Haye1}.
Beside their mathematical relevance these boundary conditions
are also of physical interest since they appear naturally
in a special one-dimensional reaction-diffusion process
with open ends \cite{Chemie}.
\\
\indent
The attempts to introduce diagonalizible
generalizations of the $XY$ model
go back to Suzuki~\cite{Suzuki}. Following these ideas
we consider generalized quantum chains
with a higher number of degrees of freedom per site.
We maintain the quantum group symmetry
by choosing special boundary conditions and
imposing appropriate restrictions on the parameters. We are interested
in both the physical properties of these generalized chains (like their
spectra) and the mathematical structure of the corresponding
quantum algebra.
\\
\indent
Let us briefly summarize the results of Ref. \cite{Haye1}.
The $XY$ chain Hamiltonian (\ref{xy}) is invariant under a
two-parameter quantum Clifford algebra which is
defined by the generators $T^1$, $T^2$ and the central element $E$
with the commutation relations
\begin{eqnarray}
\label{TwoParamAlgebra}
 & \{T^{1},T^{1}\} \;=\; 2\:[E]_{\alpha_1}\;,
\ \ \ \ \ \ \ \
 & \{T^{2},T^{2}\} \;=\; 2\:[E]_{\alpha_2} \\
 & \{T^{1},T^{2}\} \;=\; 0\;,
\ \ \ \ \ \ \ \ \ \ \ \ \ \ \
 & [E,T^{1}] \;=\; [E,T^{2}] \;=\; 0\,, \nonumber
\end{eqnarray}
\noindent
where $\alpha_1$ and $\alpha_2$ are deformation
parameters and
\begin{equation}
\label{Analog}
[E]_{\alpha_\mu} \;=\; \frac{\alpha_\mu^{E}-\alpha_\mu^{-E}}
{\alpha_\mu-\alpha_\mu^{-1}}\;\,.
\end{equation}
\noindent
The coproducts of these generators read
\begin{eqnarray}
\label{TwoCoproduct}
& & \Delta(T^1) \;=\; \alpha_1^{E/2}\otimes T^1 \;+\;
T^1\otimes\alpha_1^{-E/2} \nonumber \\
& & \Delta(T^2) \;=\; \alpha_2^{E/2}  \otimes T^2 \;+\;
T^2 \otimes\alpha_2^{-E/2} \\
& & \Delta(E) \;=\; E\otimes {\bf1}
\;+\; {\bf1} \otimes E\,. \nonumber
\end{eqnarray}
\noindent
For $\alpha_1 = \alpha_2 = 1$ the
quantum algebra (\ref{TwoParamAlgebra})
reduces to the (classical) Clifford algebra
\begin{equation}
\{T^{\mu},T^{\nu}\} \,=\, 2\,E\,\delta^{\mu\nu}\;,
\hspace{15mm}
[E,T^{\mu}] \,=\, 0\;.
\hspace{15mm}
(\mu,\nu=1,2)
\end{equation}
\noindent
Apart from the trivial one-dimensional representation the
algebra (\ref{TwoParamAlgebra})
has only two-dimen\-sio\-nal irreducible representations, in
particular the fermionic representation
corresponds to taking $T^1=\sigma^x, T^2=\sigma^y$, and $E=1$.\\
\indent
The explicit expressions for the generators in the case of the
$XY$ chain can be obtained from the fermionic (one-site)
representation by a multiple application
of the coproducts (\ref{TwoCoproduct}). In order to do so, let
us introduce local fermionic operators $\tau_j^1$
and $\tau_j^2$ by a Jordan-Wigner transformation
\begin{equation}
\label{jw}
\tau^{1}_j \;=\; \Bigl(\prod_{i=1}^{j-1}
\sigma_i^z\Bigr)\sigma_j^{x}\,,
\hspace{2cm}
\tau^{2}_j \;=\; \Bigl(\prod_{i=1}^{j-1}
\sigma_i^z\Bigr)\sigma_j^{y}
\end{equation}
\noindent
which obey the Clifford algebra
\begin{equation}
\label{TauCommutators}
\{\tau_{i}^{\mu},\tau_{j}^{\nu}\} \;=\;
2\ \delta_{ij}\delta^{\mu\nu} \;\,.
\hspace{3cm}
(i,j=1,\ldots,L;\;\;\mu,\nu=1,2)
\end{equation}
\noindent
In terms of these operators the
Hamiltonian (\ref{xy}) can be written as
\begin{equation}
\label{TauHamiltonian}
H^{XY}(\eta,q)\;=\; \frac{i}{2}\sum_{j=1}^{L-1}
\left(
   \eta\,\tau_{j}^2\tau_{j+1}^1
   \ - \  \eta^{-1}\,\tau_{j}^1\tau_{j+1}^2
   \ + \ q\tau_{j}^1\tau_{j}^2
   \ + \ q^{-1}\tau_{j+1}^1\tau_{j+1}^2 \,
\right)\,.
\end{equation}
\noindent
The explicit expressions for the
generators $T^1$, $T^2$ and $E$ read
\begin{equation}
\label{TGenerators}
T^1 \;=\;
\alpha_1^{-\frac{L+1}{2}} \;\sum_{j=1}^{L}
\alpha_1^{j} \,\tau_{j}^{1}\,, \hspace{15mm}
T^2 \;=\;
\alpha_2^{-\frac{L+1}{2}} \;\sum_{j=1}^{L}
\alpha_2^{j} \,\tau_{j}^{2}\,,  \hspace{15mm}
E\;=\;L\,,
\end{equation}
\noindent
where
\begin{equation}
\label{AlphaBeta}
\alpha_1 = \frac{q}{\eta}\,, \hspace{2cm}
\alpha_2 = q \eta\,
\end{equation}
\noindent
are the deformation parameters.
Both of them are essential, i.e. it is impossible to
remove one of the parameters by similarity transformation.
Notice furthermore that the generator
$E$ simply counts the number of sites.
Therefore if one of the deformation
parameters is a root of unity, the irreducible representations
of the algebra (\ref{TwoParamAlgebra})
depend on the length of the chain which requires to define
a special thermodynamical limit in this case~\cite{Haye1}.
\\
\indent
The generators (\ref{TGenerators}) commute with the
Hamiltonian and appear physically as a fermionic zero mode.
This zero mode is present for arbitrary parameters $q$ and $\eta$
and causes all levels of the spectrum to be
at least two-fold degenerated. We want to
emphasize that such a zero mode cannot be observed in the
case of periodic or free boundary conditions. In other words,
the quantum group symmetry is directly related to the special boundary
conditions in Eq. (\ref{xy}).
\\
\indent
If both deformation parameters $\alpha_1$ and $\alpha_2$
coincide, the total magnetization
\begin{equation}
S^z \;=\; \sum_{j=1}^L \, \sigma^z_j
\end{equation}
\noindent
also commutes with the Hamiltonian and generates an additional
$U(1)$ symmetry. In this case the quantum algebra
(\ref{TwoParamAlgebra}) can be enlarged by adding the
commutation relations
\begin{equation}
[T^1,N]=2i\,T^2\,,
\hspace{15mm}
[T^2,N]=-2i\,T^1\,,
\hspace{15mm}
[E,N]=0
\end{equation}
\noindent
and the coproduct
\begin{equation}
\Delta(N) \;=\; N\otimes {\bf1}
\;+\; {\bf1} \otimes N\,, \nonumber
\end{equation}
\noindent
where $N=\frac12(S^z+L)$.
The resulting algebra is the $U_q[SU(1/1)]$ superalgebra~\cite{Saleur}.
\\
\indent
A first attempt to generalize the quantum group invariant $XY$ chain
has been made in Ref. \cite{Lopez}. Defining a $2M$-dimensional
affine Clifford-Hopf algebra and using an $R$-matrix approach the
author showed that the generalized $XY$ chain introduced by Suzuki
\cite{Suzuki}
\begin{equation}
\label{HSuzuki}
\tilde{H} \;=\; -\sum_{k=1}^K\sum_{j=1}^{L'}\,
(\tilde{J}_{x,k} \sigma_j^x \sigma_{j+k}^x +
 \tilde{J}_{y,k} \sigma_j^y \sigma_{j+k}^y)
\sigma^z_{j+1} \ldots \sigma^z_{j+k-1} \,+\,
h\sum_{j=1}^{L'}\sigma_j^z
\end{equation}
\noindent
possesses a quantum group symmetry provided that
\begin{equation}
\tilde{J}_{x,k} \;=\; -J_x\delta_{m,k}\,,
\hspace{25mm}
\tilde{J}_{y,k} \;=\; -J_y\delta_{m,k}\,.
\hspace{25mm}
(k=1,\ldots,K)
\end{equation}
\noindent
This case is trivial for the following reason. If one performs the
transformation
\begin{eqnarray}
\label{Entwirrung}
\sigma_{mr+s}^{x,y} &\rightarrow&
(\prod_{i=0}^{r-1} \prod_{j=s+1}^{m}\, \sigma^z_{mi+j})\,
(\prod_{i=r+1}^{L'/m} \prod_{j=1}^{s-1}\, \sigma^z_{mi+j})
\,\sigma_{mr+s}^{x,y} \\
\sigma_{mr+s}^z &\rightarrow& \sigma^z_{mr+s} \nonumber
\hspace{3.5cm}
(r=0,\ldots,L'/m; \; s=1,\ldots,m)
\end{eqnarray}
\noindent
one obtains the Hamiltonian
\begin{equation}
\tilde{H}' \;=\; \sum_{r=0}^{L'/m-1} \sum_{s=1}^m\,
(\tilde{J}_x \sigma^x_{mr+s}\sigma^x_{m(r+1)+s} +
\tilde{J}_y \sigma^y_{mr+s}\sigma^y_{m(r+1)+s}) \;+\;
h\,\sum_{r=0}^{L'/m-1} \sum_{s=1}^m\,\sigma^z_{mr+s}
\end{equation}
\noindent
which is a sum of $m$ identical anisotropic $XY$ chains.
Since the transformation (\ref{Entwirrung}) does not change the
algebra of the Pauli matrices, $\tilde{H}$ and $\tilde{H}'$
differ only by a similarity transformation. Therefore the
physical properties of the $\tilde{H}$ are already known.
In this paper we show that it
is possible to define {\it nontrivial} generalizations
of the $XY$ chain maintaining both the
quantum group symmetry and the integrability in terms of free
fermions. In contrast to Ref. \cite{Lopez} we start from the physical
point of view generalizing the $XY$ chain directly in its fermionic
formulation (\ref{TauHamiltonian}). In order to implement the
quantum group symmetry we then derive a general condition for
the existence of zero modes. As an example we consider
a four-state quantum chain defined on two commuting
copies of Pauli matrices $\sigma^{x,y,z}_j$ and~$\varrho^{x,y,z}_j$.
Its Hamiltonian depends up to normalization on ten parameters:
\begin{eqnarray}
\label{FourFermionHamiltonian}
&& H\,(\gamma_1,\gamma_2,\gamma_3,\gamma_4,
\omega_{12},\omega_{34},\omega_{14},\omega_{23},
\omega_{13},\omega_{24}) \; = \; \\[2mm]
&& - \frac{1}{2}\sum_{j=1}^{L-1} \Bigl[
\omega_{12}
\;(\gamma_1^{-1} \gamma_2 \;
\sigma^x_j \varrho^z_j \sigma^x_{j+1}
\;+\;\gamma_1 \gamma_2^{-1} \; \nonumber
\sigma^y_j \varrho^z_j \sigma_{j+1}^y
\;+\;\gamma_1 \gamma_2 \;
\sigma^z_j
\;+\;\gamma_1^{-1} \gamma_2^{-1} \;
\sigma^z_{j+1}
)\\
&&\hspace{9mm}\;+\;
\omega_{34}
\;(
\gamma_3^{-1}\gamma_4 \;
\varrho_j^x \sigma_{j+1}^z \varrho^x_{j+1}
\;+\;\gamma_3\gamma_4^{-1} \; \nonumber
\varrho^y_j \sigma^z_{j+1} \varrho^y_{j+1}
\;+\; \gamma_3 \gamma_4 \;
\varrho^z_j
\;+\;\;
\gamma_3^{-1}\gamma_4^{-1}\;\varrho^z_{j+1}
)\\
&&\hspace{9mm}\;+\;
\omega_{14}
\;(
\gamma_1^{-1}\gamma_4 \;
\varrho^x_j \sigma^x_{j+1}
\;+\;\gamma_1\gamma_4^{-1} \;
\sigma^y_j \varrho^z_j \sigma^z_{j+1} \varrho^y_{j+1}
\;-\; \gamma_1 \gamma_4 \;
\sigma^y_j \varrho^y_j \nonumber
\;-\;  \gamma_1^{-1}\gamma_4^{-1}\;
\sigma^y_{j+1} \varrho^y_{j+1}
)\\
&&\hspace{9mm}\;-\;
\omega_{23}
\;(
\gamma_2^{-1}\gamma_3\;
\varrho_j^y \sigma_{j+1}^y
\;+\;\gamma_2\gamma_3^{-1} \;
\sigma^x_j \varrho^z_j \sigma^z_{j+1} \varrho^x_{j+1}
\;-\; \gamma_2 \gamma_3 \;
\sigma^x_j \varrho^x_j\nonumber
 \;-\; \gamma_2^{-1}\gamma_3^{-1} \;
\sigma^x_{j+1} \varrho^x_{j+1}
)\\
&&\hspace{9mm}\;-\;
\omega_{13}
\;(
\gamma_1^{-1}\gamma_3 \;
\varrho^y_j \sigma^x_{j+1}
\;-\;
\gamma_1\gamma_3^{-1} \;
\sigma^y_j \varrho^z_j \sigma^z_{j+1} \varrho^x_{j+1}
\;+\; \gamma_1 \gamma_3 \;
\sigma^y_j \varrho^x_j
\;+\; \gamma_1^{-1}\gamma_3^{-1} \;
\sigma^y_{j+1} \varrho^x_{j+1} \nonumber
)\\
&&\hspace{9mm}\;+\;
\omega_{24}
\;(
\gamma_2^{-1}\gamma_4 \;
\varrho_j^x \sigma_{j+1}^y \nonumber
\;-\;
\gamma_2\gamma_4^{-1} \;
\sigma^x_j \varrho^z_j \sigma^z_{j+1} \varrho^y_{j+1}
\;+\; \gamma_2 \gamma_4 \;
\sigma^x_j \varrho^y_j
\;+\; \gamma_2^{-1}\gamma_4^{-1} \;
\sigma^x_{j+1} \varrho^y_{j+1})\,  \nonumber
\Bigr]\;.
\end{eqnarray}
\noindent
Although at first sight this Hamiltonian seems to be rather artificial
we will see that it is indeed a
natural generalization of the $XY$ chain Hamiltonian (\ref{xy}).
We will show that this chain is invariant under
a four-parameter deformation of the Clifford algebra. This quantum algebra
is defined by the generators
$T^1, T^2, T^3, T^4$ and $E$ with the commutation
relations
\begin{equation}
\label{CliffordDeformed}
\{T^\mu, T^\nu \}
\;=\; 2\,\delta^{\mu\nu}\,[E]_{\alpha_\mu}\,,
\hspace{1.5cm}
[E,T^\mu] \;=\; 0\,,
\hspace{1.5cm}
(\mu,\nu=1,\ldots,4)
\end{equation}
\noindent
where $\alpha_\mu=\gamma_\mu^2$ $(\mu=1,\ldots,4)$ are
four deformation parameters. As in the case of the $XY$ chain,
we observe additional symmetries
if some of these parameters coincide.
\\
\indent
The model defined in Eq. (\ref{FourFermionHamiltonian}) can be
understood as a system of two interacting $XY$ chains:
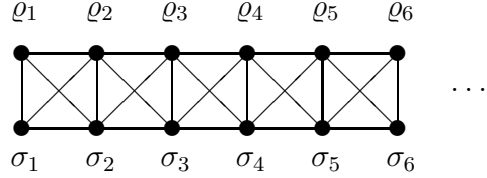
\begin{center}
\unitlength1cm
\begin{picture}(7,2)
\put(0.85,0){$\sigma_1$}
\put(1.85,0){$\sigma_2$}
\put(2.85,0){$\sigma_3$}
\put(3.85,0){$\sigma_4$}
\put(4.85,0){$\sigma_5$}
\put(5.85,0){$\sigma_6$}
\put(0.85,2){$\varrho_1$}
\put(1.85,2){$\varrho_2$}
\put(2.85,2){$\varrho_3$}
\put(3.85,2){$\varrho_4$}
\put(4.85,2){$\varrho_5$}
\put(5.85,2){$\varrho_6$}
\put(6.7,1){$\ldots$}
\multiput(1,0.5)(1,0){6}{\circle*{0.2}}
\multiput(1,1.5)(1,0){6}{\circle*{0.2}}
\multiput(1,0.5)(1,0){6}{\line(0,1){1}}
\multiput(1,0.5)(1,0){5}{\line(1,1){1}}
\multiput(1,1.5)(1,0){5}{\line(1,-1){1}}
\thicklines
\multiput(1,0.5)(1,0){5}{\line(1,0){1}}
\multiput(1,1.5)(1,0){5}{\line(1,0){1}}
\end{picture}
\end{center}
In opposition to a single $XY$ chain (which is completely described by
the deformation parameters) its Hamiltonian
depends on further six parameters~$\omega_{ij}$
which do not occur in the quantum algebra and in cannot generally be
eliminated by similarity transformation. Since these parameters
allow to implement nontrivial couplings between the
$XY$ chains without breaking the quantum group symmetry,
we expect a richer structure than in the case of decoupled
$XY$ chains as in Eq. (\ref{HSuzuki}). Switching these
couplings off by taking
$\omega_{13}=\omega_{14}=\omega_{23}=\omega_{24}=0$ and
performing the following automorphism on the Pauli matrices
\begin{eqnarray}
&&\sigma_{j}^{x,y}\;\rightarrow\;(\prod_{i=1}^{j-1}
  \varrho_{i}^{z})\,\sigma_{j}^{x,y}
\;,\hspace{12mm}
\varrho_{j}^{x,y}\;\rightarrow\;(\prod_{i=j+1}^{L}
  \sigma_{i}^{z})\,\varrho_{j}^{x,y}
\hspace{17mm}
(j=1 \ldots L)\\
&&\sigma^z_j \;\rightarrow\; \sigma^z_j\,, \hspace{27mm}
\varrho^z_j \;\rightarrow\; \varrho^z_j\, \nonumber
\end{eqnarray}
\noindent
the Hamiltonian (\ref{FourFermionHamiltonian})
decouples into a sum of two independent anisotropic
$XY$ chains:
\begin{eqnarray}
\label{Decomp}
&&H(\gamma_1, \gamma_2, \gamma_3, \gamma_4,
  \omega_{12},\omega_{34}) \; = \; \\
&&- \frac{1}{2}\sum_{j=1}^{L-1} \Bigl[\nonumber
\omega_{12}\;
(\gamma_1^{-1}\gamma_2 \;
\sigma^x_j\sigma^x_{j+1} \;+\;
\gamma_1\gamma_2^{-1} \;
\sigma^y_j \sigma_{j+1}^y \;+\;
\gamma_1 \gamma_2 \;
\sigma^z_j \;+\;
\gamma_1^{-1}\gamma_2^{-1} \;
\sigma^z_{j+1} ) \\
&&\hspace{9mm}\;+\;
\omega_{34}\;
(\gamma_3^{-1}\gamma_4 \; \nonumber
\varrho_j^x\varrho^x_{j+1} \;+\;
\gamma_3\gamma_4^{-1} \;
\varrho^y_j\varrho^y_{j+1} \;+\;
\gamma_3 \gamma_4 \;
\varrho^z_j \;+\;
\gamma_3^{-1}\gamma_4^{-1}\;
\varrho^z_{j+1} )
\,\Bigr]\,,
\end{eqnarray}
where $\omega_{12}$ and $\omega_{34}$ appear as
normalization constants.
\\
\indent
As an application we finally consider the Hamiltonian
(\ref{FourFermionHamiltonian}) for a particular choice of the
parameters $\omega_{ij}$ so that the strength of the couplings
between the two $XY$ chains is controlled by a single parameter $\xi$.
Computing the corresponding spectrum we observe that for a special
value of $\xi$ the interaction becomes singular so that
one obtains $2^{L+1}$-fold (instead of $4$-fold) degenerations.
The supplementary symmetry is caused by $L-1$ additional zero modes.
Normally zero modes are known to be
exponential modes acting globally on the whole chain.
Contrarily the additional zero modes turn out to act only
in a specific part of the chain. We thereby find a new type
of zero modes which cannot be observed in the case of
two-state models.
\\[5mm]
%
%
%
%
The paper is organized as follows. In Sect. 2 we define the class
of quantum chains to be investigated and outline the
diagonalization method. In Sect. 3 we derive
a general condition for the existence of fermionic zero modes.
Sect. 4 discusses the structure
of the corresponding quantum algebra
which is a multi-parameter deformation of
the Clifford algebra. It is shown that if some
of the deformation parameters coincide,
additional algebra automorphisms allow the number of free parameters
to be reduced. In Sect. 5 we turn our attention to a
particular physical four-state model. We discuss our results and
consider a special case where additional zero modes occur.
Finally we summarize our conclusions in Sect. 6.
Appendix A explains how the zero mode condition has been derived.
In Appendix B we show that a recently discovered
duality property of the anisotropic $XY$ chain
\cite{Haye2} also exists in the generalized case.
%
%
%

\section{Multifermionic chains and their diagonalization}
\setcounter{equation}{0}

In order to define a natural generalization of the
$XY$ chain, we first rewrite the fermionic version of the
Hamiltonian (\ref{TwoParameterHamiltonian}) in the
general bilinear form
\begin{equation}
\label{GeneralHamiltonian}
H(A,B,C) \;=\; \frac{i}{2}\;\sum_{j=1}^{L-1}\,
\sum_{\mu,\nu=1}^{2n}
\left(A^{\mu,\nu}\tau_{j}^{\mu}\tau_{j+1}^{\nu}
\;+\;\frac{1}{2}B^{\mu,\nu}\tau_{j}^{\mu}\tau_{j}^{\nu}
\;+\;\frac{1}{2}C^{\mu,\nu}
\tau_{j+1}^{\mu}\tau_{j+1}^{\nu}\right)\,,
\end{equation}
\noindent
where $n=1$ and
\begin{equation}
\label{ABC}
A \,=\, \left(
    \begin{array}{cc}
    0 & -\eta^{-1} \\
    \eta & 0
    \end{array}
\right),
\;\;\;\;\;\;\; B \,=\, \left(
    \begin{array}{cc}
 	   0 & q \\
    -q & 0
    \end{array}
\right),
\;\;\;\;\;\;\; C \,=\, \left(
    \begin{array}{cc}
    0 & q^{-1} \\
    -q^{-1} & 0
    \end{array}
\right)\,.
\end{equation}
\noindent
It is near at hand to consider the
Hamiltonian (\ref{GeneralHamiltonian})
for $n>1$ pairs of local fermionic operators $\tau_j^\mu$ per site.
These operators are supposed to obey
the Clifford algebra
\begin{equation}
\label{MultiFermCliff}
\{\tau_{i}^{\mu},\tau_{j}^{\nu}\} \;=\;
2\ \delta_{ij}\delta^{\mu\nu} \;
\hspace{2cm}
(i,j=1 \ldots L;\;\;\mu,\nu = 1, \ldots, 2n)
\end{equation}
and therefore the $2n \times 2n$
matrices $B$ and $C$
can be assumed to be antisymmetric (since symmetric
contributions would result in an irrelevant constant).
This defines a $2n$-state model which is a
natural generalization of the $XY$ chain
in the fermionic language. As we will see below,
the four-state Hamiltonian (\ref{FourFermionHamiltonian}) just
corresponds to the case $n=2$.
\\
\indent
We are now going to diagonalize these generalized quantum chains.
Since the Hamiltonian~(\ref{GeneralHamiltonian}) is bilinear
in the operators $\tau_j^\mu$,
it is possible to apply standard methods described in Ref. \cite{LSM}.
Accordingly $H$ can be written in the diagonal form
\begin{equation}
\label{DiagonalForm}
H(A,B,C) \;=\;\sum_{k=0}^{L-1}\,
      \sum_{\gamma=1}^n
      \Lambda_{k}^{\gamma}\;
      i T_{k}^{2\gamma} \, T_{k}^{2\gamma-1} \;
\end{equation}
\noindent
where $\Lambda_k^\gamma$ are fermionic excitation energies and
$T_k^\mu$ are diagonal Clifford operators:
\begin{equation}
\label{TCommutators}
\{T_{k}^{\mu},T_{l}^{\nu}\} \;=\;
2\ \delta_{kl}\delta^{\mu\nu}\,.
\;\;\;\;\;\;\;\;\;\;\;\;\;\;\;\;\;\;\;\;\;\;
(k,l=0,\ldots,L-1;\;\;\mu,\nu=1,\ldots,2n)
\end{equation}
\noindent
They are related to the local fermionic operators $\tau_j^\mu$
by an orthogonal transformation:
\begin{equation}
\label{TDefinition}
T_{k}^{\nu}\;=\;\sum_{j=1}^{L}\,\sum_{\mu=1}^{2n}\,
\psi_{k,j}^{\nu,\mu}\,\tau_{j}^{\mu}\;,
\hspace{20mm}
\sum_{j=1}^L\,\sum_{\mu=1}^{2n}\,
\psi_{k,j}^{\gamma,\mu}\,\psi_{l,j}^{\delta,\mu}\;=\;
\delta_{kl}\delta^{\gamma\delta}
\end{equation}
\noindent
Thus the expressions
$i T_{k}^{2\gamma} \, T_{k}^{2\gamma-1} $ in Eq. (\ref{DiagonalForm})
have the eigenvalues $\pm 1$ so that the knowledge of all
excitation energies $\Lambda_k^\gamma$ allows the spectrum of
the Hamiltonian (\ref{GeneralHamiltonian}) to be constructed by
taking all combinations into account.
As shown in Ref. \cite{LSM} the excitation energies $\Lambda_k^\gamma$
and the transformation coefficients $\psi_{k,j}^{\gamma,\mu}$ are
solutions of the eigenvalue problem
\begin{equation}
\label{EVP}
\sum_{j,\nu}\,M_{ij}^{\mu\nu}\,\Phi_{kj}^{\gamma\nu} \;=\;
\mp 2i\,\Lambda_k^\gamma \,\Phi_{ki}^{\gamma\mu}\,,
\hspace{2cm}
(\gamma=1,\ldots,n; \;\;\; \mu=1,\ldots,2n)
\end{equation}
\noindent
where $\Phi_{ki}^{\gamma\mu} = \psi_{k,i}^{2\gamma-1,\mu} \pm
\psi_{k,i}^{2\gamma,\mu}$ and $M$ is the following
$2nL \times 2nL$ matrix:
\begin{equation}
\label{M}
M \,=\, \left(
    \begin{array}{cccccc}
    B & A & & & & \\
    -A^{T} & B+C & A & & & \\
    & -A^{T} & B+C & A & & \\
    & & ... & ... & ... & \\
    & & & -A^{T} & B+C & A \\
    & & & & -A^{T} & C
    \end{array}
\right)\;.
\end{equation}
\noindent
Since $M$ is antisymmetric we expect its eigenvalues to occur
in pairs with different signs. We thus are free to choose the
sign of $\Lambda_k^\gamma$. However, the spectrum
does not depend on this choice.
\\
\indent
Let us summarize our results at this stage. We have constructed
a class of Hamiltonians of the form (\ref{GeneralHamiltonian})
defined on $n$ pairs of fermionic operators per site.
These chains depend up to normalization
on $2n^2-n$ parameters which are arranged in
$2n \times 2n$ matrices $A$, $B$ and $C$.
Their spectra can be determined by
solving the reduced eigenvalue problem~(\ref{EVP}).
In the following section we are going to derive an additional condition
on the matrices $A$, $B$ and $C$ in order to implement the quantum
group symmetry and to eliminate unessential parameters.

%
%

\section{A condition for the existence of zero modes}
\setcounter{equation}{0}

In case of the $XY$ chain the quantum group symmetry appears
as a fermionic zero mode $\Lambda_0=0$
for {\it arbitrary} parameters $\eta$ and $q$.
Our aim is to implement a similar structure
in the case of generalized chains (\ref{GeneralHamiltonian}).
We therefore search for a general condition which implies
the existence of fermionic zero modes.
According to Eq. (\ref{EVP}) such zero modes are solutions
of the system of equations
\begin{equation}
\label{ZMC}
\sum_{j=1}^L \sum_{\nu=1}^{2n} M_{ij}^{\mu\nu}
\psi_{0,j}^{\gamma,\nu} \;=\; 0\,.
\hspace{20mm}
(i=1,\ldots,L;\,\, \mu=1,\ldots,2n;\,\,\gamma=1,\ldots,n)
\end{equation}
\noindent
As a necessary condition the determinant of $M$ therefore
has to vanish. Since it is quite difficult to solve this problem
in general we are only interested in a special set of solutions.
As shown in Appendix A the systems of equations (\ref{ZMC})
simplifies essentially if the matrices
$A$, $B$, and $C$ satisfy the condition
\begin{equation}
\label{ZeroModeCondition}
A^{T}+CA^{-1}B=0\,.
\end{equation}
\noindent
This {\it zero mode condition} is assumed to be valid
throughout the rest of this paper.
It implies that the components
of the zero mode eigenvectors $\vec{\psi}_{0,j}^\mu =
(\psi_{0,j}^{\mu,1},\ldots,\psi_{0,j}^{\mu,2n})$
obey a simple power-law:
\begin{equation}
\label{PowerLaw}
\vec{\psi}_{0,j}^\nu = (-A^{-1}B)^{j-1}\,\vec{\psi}_{0,1}^\nu\,.
\end{equation}
\noindent
It is easy to check that in the $XY$ chain
case the matrices (\ref{ABC}) satisfy the zero mode condition
(\ref{ZeroModeCondition}). We now consider the generalized
case $n>1$. In order to simplify the eigenvectors~(\ref{PowerLaw})
and remove unessential parameters
one can use the invariance of the Clifford
algebra~(\ref{TauCommutators}) under orthogonal transformations $O(2n)$:
\begin{equation}
\label{OrthoTransform}
\tau_i^\mu \;\rightarrow\;
{\tau_i^\mu}' =
\sum_{\nu=1}^{2n} O^{\mu\nu} \,
\tau_i^\nu \,.
\hspace{2cm}
(OO^T=O^TO = \id)\,
\end{equation}
\noindent
Therefore a change of basis
\begin{eqnarray}
\label{ABCTransform}
A & \rightarrow & A'=O\,A\,O^T \nonumber \\
B & \rightarrow & B'=O\,B\,O^T  \\
C & \rightarrow & C'=O\,C\,O^T \nonumber \,
\end{eqnarray}
\noindent
corresponds to a similarity transformation of the
Hamiltonian (\ref{GeneralHamiltonian}):
\begin{equation}
\label{ahnlich}
H(A',B',C') = U\,H(A,B,C)\,U^{-1}\,.
\end{equation}
This allows us to choose a basis where the matrix $-A^{-1}B$
in Eq. (\ref{PowerLaw}) is already diagonal:
\begin{equation}
\label{SpecChoice}
(-A^{-1}B)^{\mu\nu} \;=\; \alpha_\mu\,\delta^{\mu\nu}\,.
\end{equation}
\noindent
According to Eq. (\ref{TDefinition}) the zero mode
operators then read
\begin{equation}
\label{MultiFermZeroMode}
T_0^\mu \;=\; (\alpha_\mu)^{-\frac{L+1}2}
\sum_{j=1}^L\,(\alpha_\mu)^j\,\tau_j^\mu\,
\hspace{2cm}
(\mu =1, \ldots, 2n)
\end{equation}
Because of $\Lambda_0^\gamma=0$ (c.f. Eq. (\ref{DiagonalForm}))
these operators commute with $H(A,B,C)$ and therefore all levels of
the spectrum are at least $2^n$-fold degenerated.
As will be seen in the next section, they appear as the generators
of the corresponding quantum algebra.
\\
\indent
Another very useful advantage of the zero mode condition
(\ref{ZeroModeCondition}) is a further simplification of the
eigenvalue problem (\ref{EVP}). It turns out that the
eigenvalues of $M$ (beside the zero modes $\Lambda_0^\mu=0$)
are the solutions of the polynomial
\begin{equation}
\label{DispCharPol}
\det \Bigl(
-A^T\,e^{-i \pi k/L} \;+\; (B+C-2i\Lambda_k^\gamma) \;+\;
A\,e^{i \pi k/L}\, \Bigr)
\;=\; 0\,,
\end{equation}
\noindent
where $k$ runs from $1$ to $L-1$.
This polynomial contains only even powers of $\Lambda_k^\gamma$
(due to the freedom of choosing its sign) and constitutes the
dispersion relation of the chain.
\\
\indent
Notice that the zero modes are always related to exponential wave
functions and cannot be derived from Eq. (\ref{DispCharPol}).
Here it is useful to give some comment.
It is a well-known property of integrable
quantum chains with open boundary conditions that beside the
excitations with trigonometric wave functions
there is always a set of exceptional excitations with
exponential behaviour. In the thermodynamic limit
$L \rightarrow \infty$ these wave functions  are
located at the ends of the chain and have a vanishing energy.
In our models a special choice of the boundary
conditions causes these excitation energies
to vanish exactly {\it for finite} $L$ giving the exponential
wave functions the physical meaning of zero modes.
\\
\indent
Hamiltonians of the form (\ref{GeneralHamiltonian})
obeying the zero mode condition (\ref{ZeroModeCondition})
in the basis (\ref{SpecChoice})
can be constructed as follows.
Let us choose an arbitrary diagonal
$2n \times 2n$ matrix $\Gamma$ and an arbitrary antisymmetric
$2n \times 2n$ matrix $\Omega$. Then the matrices
\begin{equation}
\label{SpecialChoice}
A=-\Gamma\Omega\Gamma^{-1}\,, \hspace{15mm}
B=\Gamma\Omega\Gamma\,, \hspace{15mm}
C=\Gamma^{-1}\Omega\Gamma^{-1} \,
\end{equation}
\noindent
satisfy the zero mode condition (\ref{ZeroModeCondition})
and $-A^{-1}B=\Gamma^2$ is already diagonal. The corresponding
Hamiltonian $H(\Omega,\Gamma)$ therefore depends
up to normalization on $2n^2+n$ parameters.
\\
\indent
Let us illustrate this construction for the case $n=2$.
Choosing
\begin{equation}
\label{omega}
\Omega \,=\, \left(
    \begin{array}{cccc}
    0 & \omega_{12} & \omega_{13} & \omega_{14}  \\
    -\omega_{12} & 0 & \omega_{23} & \omega_{24} \\
    -\omega_{13} & -\omega_{23} & 0 & \omega_{34} \\
    -\omega_{14} & -\omega_{24} & -\omega_{34} & 0 \end{array}
\right)\,, \hspace{2cm}
\Gamma \,=\, \left(
    \begin{array}{cccc}
    \gamma_1 &&&  \\
    & \gamma_2 &&  \\
    && \gamma_3 &  \\
    &&& \gamma_4 \end{array}
\right)\,
\end{equation}
and inserting the matrices (\ref{SpecialChoice})
into Eq. (\ref{GeneralHamiltonian}) we obtain a ten-parameter
Hamiltonian with two fermionic zero modes.
Their deformation parameters $\alpha_\mu$ in Eq.
(\ref{MultiFermZeroMode}) are
simply given by $\alpha_\mu=\gamma_\mu^2$.
Then performing a generalized Jordan-Wigner transformation
\begin{eqnarray}
\label{jw2}
&&\tau^{1}_j \;=\; \Bigl(\prod_{i=1}^{j-1}
\sigma_i^z\rho_i^z\Bigr)\sigma_j^{x}
\hspace{16mm}
\tau^{2}_j \;=\; \Bigl(\prod_{i=1}^{j-1}
\sigma_i^z\rho_i^z\Bigr)\sigma_j^{y}
\\
&&\tau^{3}_j \;=\; \Bigl(\prod_{i=1}^{j-1}
\sigma_i^z\rho_i^z\Bigr)\sigma_j^{z}\rho_j^x
\hspace{13mm}
\tau^{4}_j \;=\; \Bigl(\prod_{i=1}^{j-1}
\sigma_i^z\rho_i^z\Bigr)\sigma_j^{z}\rho_j^y
\nonumber
\end{eqnarray}
\noindent
one obtains directly the ten-parameter Hamiltonian
(\ref{GeneralHamiltonian}). It is now clear that the
somewhat artificial appearance of this Hamiltonian is
nothing but a simple consequence of Jordan-Wigner factors while in the
fermionic formulation the generalization is
a quite natural one.
\\
\indent
We now apply Eq. (\ref{DispCharPol}) in order to compute the
spectrum of the Hamiltonian (\ref{GeneralHamiltonian}).
One obtains the fermionic excitation energies
\begin{eqnarray}
\label{FourFermionExcitations}
&&
\Lambda_k^{1}\;=\;\sqrt{p_k + \sqrt{p_k^2-q_k}}\,,
\hspace{3cm}
(k=1,\ldots,L-1) \\
&&
\Lambda_k^{2}\;=\;\sqrt{p_k - \sqrt{p_k^2-q_k}}\,,
\end{eqnarray}
\noindent
where
\begin{eqnarray}
\label{explicit}
p_k &=& \frac12\,
\sum_{1 \leq \mu < \nu \leq 4}
(\cos{\frac{\pi k}L}-\frac{\alpha_\mu+\alpha_\mu^{-1}}{2})\,
(\cos{\frac{\pi k}L}-\frac{\alpha_\nu+\alpha_\nu^{-1}}{2})
\,\omega_{\mu\nu}^2
\\
q_k &=& (\omega_{12}\omega_{34}-\omega_{13}\omega_{24}+
         \omega_{14}\omega_{23})^2\,
\,\prod_{\mu=1}^4 \,(\cos{\frac{\pi k}L}-
\frac{\alpha_\mu+\alpha_\mu^{-1}}{2})\,.
\end{eqnarray}
\noindent
The levels of the spectrum can be computed by
taking all fermionic combinations
into account (see Eq. (\ref{DiagonalForm})).
Because of the zero modes $\Lambda_0^1=\Lambda_0^2=0$
each level is at least four-fold degenerated.
Obviously the spectrum is massless if at least one
of the deformation parameters is on the unit circle.
Moreover we observe that the
spectrum is invariant under
discrete transformations $\alpha_\mu \rightarrow \alpha^{-1}_\mu$.
This symmetry is related to a generalized duality
property which will be discussed in Appendix B.
%
%
%
%
\section{The Clifford quantum algebra}
\setcounter{equation}{0}
If the Hamiltonian (\ref{GeneralHamiltonian}) satisfies
the zero mode condition (\ref{ZeroModeCondition}), it
is invariant under a $2n$-parameter deformation of the
Clifford algebra. This quantum algebra is defined
by the commutation relations
\begin{equation}
\label{DefCliff}
\{T^\mu, T^\nu \}
\;=\; 2\,\delta^{\mu\nu}\,[E]_{\alpha_\mu}\,,
\hspace{1.5cm}
[E,T^\mu] \;=\; 0
\hspace{1.5cm}
(\mu,\nu=1,\ldots,2n)
\end{equation}
\noindent
and the coproducts
\begin{eqnarray}
\label{MultiCoproduct1}
\Delta(T^\mu) &=& \alpha_\mu^{E/2}\otimes T^\mu \;+\;
T^\mu \otimes \alpha_\mu^{-E/2}
\hspace{2cm} (\mu=1,\ldots,2n) \\
\label{MultiCoproduct2}
\Delta(E) &=& E\otimes {\bf1}
\;+\; {\bf1} \otimes E\,
\hspace{3.5cm}
(\mu,\nu=1,\ldots,2n)
\end{eqnarray}
\noindent
with the co-unit
\begin{equation}
\epsilon(T^\mu)\;=\;\epsilon(E)\;=\;0
\end{equation}
\noindent
and the antipode
\begin{equation}
S(T^\mu)\;=\;T^\mu\,,
\hspace{22mm}
S(E)\;=\;-E\,.
\end{equation}
\noindent
It has been given in a similar form in Ref. \cite{Spanier},
where $2n$ distinct central elements and one deformation parameter
have been used (instead of $2n$ deformation parameters and one
central element $E$ in our case which leads to a
different representation theory).
Notice that by construction of our model the dimension of
the algebra (\ref{DefCliff}) is always even (the odd case
however is also possible but not of interest in this paper).
If all deformation parameters
$\alpha_1,\ldots,\alpha_{2n}$ are equal to one, the algebra
reduces to the (classical) Clifford algebra
\begin{equation}
\{T^\mu, T^\nu \}
\;=\; 2\,\delta^{\mu\nu}\,.
\end{equation}
\noindent
Beside the trivial one-dimensional representation
$T^\mu=E=0$ the algebra (\ref{DefCliff}) possesses only
$2n$-dimensional irreducible representations of the form
\begin{equation}
\label{TwoIrrep}
T^\mu \;=\; \sqrt{[e]_{\alpha_\mu}}\, t^\mu\,,
\hspace{2cm}
E\;=\; e\,{\bf 1}\,,
\end{equation}
\noindent
where $e$ is a number and the $t^\mu$ denote a canonical
representation of the $2n$-dimensional
classical Clifford algebra
$\{t^\mu,t^\nu\}=2\delta^{\mu\nu}$. For $n=2$ a
possible choice is
\begin{equation}
t^{+1} = \sigma^x \otimes {\bf 1} \hspace{1cm}
t^{-1} = \sigma^y \otimes {\bf 1} \hspace{1cm}
t^{+2} = \sigma^z \otimes \sigma^x \hspace{1cm}
t^{-2} = \sigma^z \otimes \sigma^y\,.
\end{equation}
\noindent
In particular the fermionic representation corresponds to
taking $e=1$. The coproduct (\ref{MultiCoproduct1}) then
explicitely reads
\begin{equation}
\Delta(t^\mu)\;=\;
\alpha_\mu^{1/2}\,t^{2n+1} \otimes t^\mu\,+\,
\alpha_\mu^{-1/2}\,t^\mu \otimes {\bf 1}\,,
\end{equation}
\noindent
where $t^{2n+1}=\sigma^z\otimes\sigma^z$ plays the
role of a grading operator.
By a multiple application of this coproduct
we  obtain the $L$-site representation
\begin{equation}
T^\mu \;=\;
\sum_{j=1}^L\,(\alpha_\mu)^{j-\frac{L+1}2}\,\tau_j^\mu\,,
\hspace{12mm}
E\;=\;L\,.
\hspace{15mm}
(\mu =1, \ldots, 2n)
\end{equation}
\noindent
These generators are nothing but the zero mode operators defined
in Eq. (\ref{MultiFermZeroMode}). They commute
with the Hamiltonian (\ref{GeneralHamiltonian}) and
therefore the chain is invariant under the deformed
Clifford algebra (\ref{DefCliff}).
\\
\indent
If one  of the deformation
parameters $\alpha_1,\ldots,\alpha_{2n}$ is a root of unity
(i.e. the parameters are non-generic)
the r.h.s. of Eq. (\ref{DefCliff}) may vanish. In this case
the two-dimensional irreducible representations (\ref{TwoIrrep})
break down and only the trivial one survives. In the spectrum
non-generic cases appear as level crossings. Here the Hamiltonian
possesses zero-norm eigenvectors and one has to consider an appropriate
subspace and a redefined scalar product. We do not want to discuss this
case in detail here and therefore we will assume the deformation
parameters to be generic.
\\
\indent
If some of the deformation parameters $\alpha_1,\ldots,\alpha_{2n}$
coincide, it is possible to perform orthogonal transformations
(\ref{OrthoTransform}) in the corresponding subspace
{\it without} altering the commutation relations (\ref{DefCliff})
This allows further parameters in the matrix $\Omega$
(see Eq. (\ref{omega})) to be eliminated.
Let us first consider the case where
all deformation parameters coincide. Then by means of $O(2n)$
rotations we can always transform the matrix $\Omega$ to the
block-diagonal form
\begin{equation}
\label{QuasiDiag}
\left(
    \begin{array}{cccccc}
    0 & \omega_{1,2} & & & & \\
    -\omega_{1,2} & 0 & & & & \\
    & & ... & & & \\
    & & & ... & & \\
    & & & & 0 & \omega_{2n-1,2n} \\
    & & & & -\omega_{2n-1,2n} & 0
    \end{array}
\right)\,.
\end{equation}
Hence for equal deformation parameters the generalized
Hamiltonian (\ref{GeneralHamiltonian}) (together with the
zero mode condition (\ref{ZeroModeCondition})) always decomposes
into a sum of $n$ isotropic $XY$ chains of the form (\ref{xy}).
However, the normalizations and deformation parameters $q$
of each copy may be different. \\
\indent
If only $m<2n$ deformation parameters coincide (while the others
are pairwise distinct), similar considerations show that one can
remove $\frac12(m^2-m)$ parameters in the matrix $\Omega$ by
means of $O(m)$ transformations.

%
%
%
%

\section{A special physical example}
In this section we want to illustrate our results
in the example of the chain (\ref{FourFermionHamiltonian}).
We only consider a special choice of the
coupling constants $\omega_{ij}$ to be defined below. This choice
is motivated by physical reasons as follows. Thinking of two coupled
$XY$ chains and neglecting the influence
of the deformation parameters
we suppose the internal interactions of each chain
to have the same strength ($\omega_{12}=\omega_{34}=1$).
On the other hand there are couplings between
both chains which can be controlled by a
single parameter $\xi$.
Since within each $XY$ chain only $X$-$X$
and $Y$-$Y$ interactions are present, we assume for physical reasons
the same to be true for the interactions between both chains, i.e.
we exclude $X$-$Y$ interactions by setting $\omega_{13}=\omega_{24}=0$.
Therefore taking care of the signs in Eq. (\ref{FourFermionHamiltonian})
a physically reasonable choice of the coupling constant is
\begin{equation}
\omega_{12} = \omega_{34} = 1\,, \hspace{13mm}
\omega_{14} = -\omega_{23} = \xi\,, \hspace{13mm}
\omega_{13} = \omega_{24} = 0 \,.
\end{equation}
\noindent
The Hamiltonian (\ref{FourFermionHamiltonian}) then reads
\begin{eqnarray}
\label{Hquer}
&& H\,(\gamma_1,\gamma_2,\gamma_3,\gamma_4,\xi) \; = \; \\[2mm]
&& - \frac{1}{2}\sum_{j=1}^{L-1} \Bigl[ \hspace{5mm}
\gamma_1^{-1}\gamma_2 \;
\sigma^x_j \varrho^z_j \sigma^x_{j+1}
\;+\;\gamma_1\gamma_2^{-1} \; \nonumber
\sigma^y_j \varrho^z_j \sigma_{j+1}^y
\;+\;\gamma_1 \gamma_2 \;
\sigma^z_j
\;+\;\gamma_1^{-1}\gamma_2^{-1} \;
\sigma^z_{j+1}\\[1mm]
&&\hspace{9mm}\;+\; \hspace{5mm}
\gamma_3^{-1}\gamma_4 \;
\varrho_j^x \sigma_{j+1}^z \varrho^x_{j+1}
\;+\;\gamma_3\gamma_4^{-1} \; \nonumber
\varrho^y_j \sigma^z_{j+1} \varrho^y_{j+1}
\;+\; \gamma_3 \gamma_4 \;
\varrho^z_j
\;+\; \gamma_3^{-1}\gamma_4^{-1}\;
\varrho^z_{j+1}\\[1mm]
&&\hspace{9mm}\;+\;
\xi\,(\,\gamma_1^{-1}\gamma_4 \;
\varrho^x_j \sigma^x_{j+1}
\;+\;\gamma_1\gamma_4^{-1} \;
\sigma^y_j \varrho^z_j \sigma^z_{j+1} \varrho^y_{j+1}
\;-\; \gamma_1 \gamma_4 \;
\sigma^y_j \varrho^y_j \nonumber
\;-\; \gamma_1^{-1}\gamma_4^{-1} \;
\sigma^y_{j+1} \varrho^y_{j+1}\,)\\
&&\hspace{9mm}\;+\;\xi\,(
\gamma_2^{-1}\gamma_3\;
\varrho_j^y \sigma_{j+1}^y
\;+\;\gamma_2\gamma_3^{-1} \;
\sigma^x_j \varrho^z_j \sigma^z_{j+1} \varrho^x_{j+1}
\;-\; \gamma_2 \gamma_3 \;
\sigma^x_j \varrho^x_j\nonumber
\;-\; \gamma_2^{-1}\gamma_3^{-1} \;
\sigma^x_{j+1} \varrho^x_{j+1}\,)\,
\Bigr]\;.
\end{eqnarray}
\noindent
Let us first consider the case where all deformations
$\gamma_1=\gamma_2=\gamma_3=\gamma_4=\gamma$ are equal.
As shown in Fig. 1,
the spectrum of $H$ depends linearly on $\xi$ in this case.
For $\xi=0$ the spectrum is just the sum of two identical
isotropic $XY$ chains
\begin{equation}
H(\gamma,\gamma,\gamma,\gamma,0)
\;\doteq\; H^{XY}(1,\gamma^2) \otimes \id\,+\,
\id \otimes H^{XY}(1,\gamma^2) ,
\end{equation}
\noindent
where $H^{XY}(\eta,q)$ is given in Eq. (\ref{xy}) and `$\doteq$' denotes
equality up to similarity transformation.
Looking at Fig. 1 it is obvious that for arbitrary $\xi$
only the normalizations of the two $XY$ chains vary linearly:
\begin{equation}
H(\gamma,\gamma,\gamma,\gamma,\xi)
\;\doteq\; (1+\xi)\,H^{XY}(1,\gamma^2)  \otimes \id\,+\,
(1-\xi)\,\id \otimes H^{XY}(1,\gamma^2)
\end{equation}
\noindent
Thus for equal deformation parameters the generalized chain
(\ref{Hquer}) always decouples into a sum of two isotropic
$XY$ chains in agreement with the observation in Eq. (\ref{QuasiDiag}).
In particular for $\xi=1$ only one of them
survives so that each level of the spectrum is at
least $2^{L+1}$-fold degenerated. Here the Hamiltonian (\ref{Hquer})
is invariant under local rotations generated
by the anticommuting operators
\begin{equation}
\label{LocalRotations}
S^1_j \;=\; \tau_j^1-\tau_j^3\,,
\hspace{15mm}
S^2_j \;=\; \tau_j^2-\tau_j^4\,.
\hspace{18mm}
(j=1,\ldots,L)
\end{equation}
\noindent
We will come back to this case below.
\\
\indent
For arbitrary deformation parameters the spectrum
cannot be decomposed into a sum of $XY$ chain spectra (see Fig. 2).
The only exceptions are $\xi=0$ and $\xi=1$. In the first case
we have
\begin{equation}
\label{zeq0}
H(\gamma_1,\gamma_2,\gamma_3,\gamma_4,0)
\;\doteq\; H^{XY}(\gamma_1^{-1}\gamma_2,\gamma_1\gamma_2)
\otimes \id \,+\, \id \otimes
H^{XY}(\gamma_3^{-1}\gamma_4,\gamma_3\gamma_4)
\end{equation}
in agreement with Eq. (\ref{Decomp}). For $\xi=1$ we claim that
the spectrum of $H$ coincides up to degenerations
with the spectrum of a single anisotropic $XY$ chain:
\begin{equation}
\label{zeq1}
H(\gamma_1,\gamma_2,\gamma_3,\gamma_4,1)
\;\doteq\; 2\,H^{XY}(\eta,q) \otimes \id \,.
\end{equation}
Here the parameters $\eta$ and $q$
are solutions of the following trigonometric equations
\begin{eqnarray}
\nu &=&  { \left(\frac{\eta+\eta^{-1}}{2}\right)}^{2} +
    {      \left(\frac{q+q^{-1}}{2}\right)}^{2} -1
    \;=\; \frac{1}{4}(\Delta_1+\Delta_3)(\Delta_2+\Delta_4)\,,
\\[4mm]
\mu &=&
    4\:{\left(\frac{\eta-\eta^{-1}}{2}\right)}^{2}\;
    {   \left(\frac{q-q^{-1}}{2}\right)}^{2}
    \;=\; \frac{1}{4}(\Delta_1+\Delta_3-\Delta_2-\Delta_4)^2\,,
\end{eqnarray}
\noindent
where
\begin{equation}
\Delta_i\;=\;\frac{\alpha_i+\alpha_i^{-1}}{2} \;=\;
\frac{\gamma_i^2+\gamma_i^{-2}}{2}\,.
\end{equation}
\noindent
In order to prove Eq. (\ref{zeq1}) we checked
that the Hamiltonian
\begin{equation}
H(\gamma_1,\gamma_2,\gamma_3,\gamma_4,1) \;=\;
2 \sum_{j=1}^{L-1} h_j
\end{equation}
\noindent
satisfies indeed the algebraic relations of the anisotropic
$XY$ chain \cite{Haye1}:
\begin{eqnarray}
\label{e19}
& & \Bigl[ h_{j}h_{j\pm1}h_{j}-
h_{j\pm1}h_{j}h_{j\pm1} + (\nu-1)
(h_{j}-h_{j\pm1}) \Bigr] \:
(h_{j}-h_{j\pm1}) \;=\; \mu\; \\[2mm]
& & \hspace{4cm}
h_{j}^{2} \;=\; \nu\;,
\end{eqnarray}
\noindent
Hence for arbitrary deformation parameters
a special tuning of the coupling constants (due to the choice $\xi=1$)
provides a strong symmetry. The spectrum is
equivalent to that of a single
anisotropic XY chain and each level is at least
$2^{L+1}$-fold degenerated.
The corresponding symmetry operators read
\begin{eqnarray}
\label{LeftOps}
&&  L^1_j \;=\; \sum_{k=1}^j\,\bigl(\alpha_1^{k-j-1/2}\tau_j^1 -
			     \alpha_3^{k-j-1/2}\tau_j^3 \bigr)
\hspace{20mm} (j=1,\ldots,L-1)\\
&&  L^2_j \;=\; \sum_{k=1}^j\,\bigl(\alpha_2^{k-j-1/2}\tau_j^2 -
			     \alpha_4^{k-j-1/2}\tau_j^4 \bigr)
\nonumber \\[1mm]
&&  [L_i^\mu,H] \;=\; 0\,. \nonumber
\end{eqnarray}
\noindent
These generators may be understood as $L-1$ additional zero modes.
Together with the four zero mode generators
$T_0^\mu$ they cause $2^{L+1}$-fold degenerations.
\\
\indent
The special property of the zero modes (\ref{LeftOps})
is that they act only in a part of the chain extending
from the left boundary to a certain position $j$. Similarly there are
zero mode operators acting from position $j+1$ to the right boundary:
\begin{eqnarray}
\label{RightOps}
&&  R^1_j \;=\; \sum_{k=j+1}^L\,\bigl(\alpha_1^{k-j-1/2}\tau_j^1 -
			     \alpha_3^{k-j-1/2}\tau_j^3 \bigr)
\hspace{20mm} (j=1,\ldots,L-1)\\
&&  R^2_j \;=\; \sum_{k=j+1}^L\,\bigl(\alpha_2^{k-j-1/2}\tau_j^2 -
			     \alpha_4^{k-j-1/2}\tau_j^4 \bigr)
\nonumber
\end{eqnarray}
\noindent
Because of
\begin{equation}
L_j^\mu+R_j^\mu \;=\; \alpha_\mu^{L/2-j}\,T_0^\mu \,-\,
		      \alpha_{\mu+2}^{L/2-j}\,T_0^{\mu+2}
\hspace{20mm} (\mu=1,2)
\end{equation}
\noindent
only one set of operators (e.g. $\{L_j^\mu\}$) is independent.
It is easy to check that for all
deformation parameters being equal
one retrieves the local symmetry
operators (\ref{LocalRotations}) by taking
appropriate linear combinations.
\\ \indent
As already mentioned in Sect. 3, the existence of exponential modes
is a well-known property of integrable chains with non-periodic
boundary conditions. Normally there are only four exponential modes
in our model,
namely the zero modes (\ref{MultiFermZeroMode}). These modes always
act globally. In contrast for $\xi=1$
the additional zero modes (\ref{LeftOps})
and (\ref{RightOps}) act only to the left and to the right of a
certain position, respectively. To our knowledge this phenomenon
has not been observed before. It has its origin in a singularity of
the interaction (det$(\Omega)=0$) for $\xi=1$. Roughly speaking
certain modifications of the states at site $j$
do not affect the situation at site $j+1$. Therefore
if one combines exponential modes in a appropriate way they
`trickle away' at a certain position.
%
%
%
%
\section{Conclusions}
\setcounter{equation}{0}
The present work is based on previous investigations of the anisotropic
$XY$ chain in a magnetic field with a special kind of boundary
conditions. These boundary conditions imply the existence of a
fermionic zero mode which is related to a quantum group symmetry.
\\
\indent
In this article we found a class of integrable quantum chains
which can be understood as generalizations of the $XY$ chain.
These $2n$-state models are defined on $n$ fermionic degrees of freedom
per site and can be diagonalized in terms of free fermions as well.
In analogy to the $XY$ chain case we found a general condition
for the existence of fermionic zero modes.
This condition implies in turn that the Hamiltonian is invariant
under a $2n$-parameter deformation of the $2n$-dimensional Clifford
algebra and causes $2^n$-fold degenerations of each energy level.
Discussing the structure of this algebra we observed that if some
of the deformation parameters coincide, the symmetry of the
chain is increased by means of orthogonal algebra automorphisms
leading to higher degenerations of the spectrum.
\\
\indent
The structure of the quantum group allows complicated internal
couplings to be implemented. These couplings are nontrivial in the
sense that the spectrum of such a chain cannot be decomposed into a
sum of $XY$ chain spectra. As an example we discussed
a four-state model and computed the corresponding spectrum.
In this case one can think of two $XY$ chains
with nearest-neighbour couplings
between them. The corresponding Hamiltonian
depends on ten parameters, four of them being deformation
parameters of the Clifford algebra. We restricted our attention to a
special choice of the other six parameters which is supposedly
the most physical one (we allow only $XX$ and $YY$ couplings
between the chains) and illustrated our results.
For a special tuning of the coupling constants
the interaction matrix becomes singular and the spectrum coincides
with that of a single $XY$ chain. However, the degenerations are
much larger due to the existence of $L-1$ additional zero modes.
In contrast to usual exponential modes
zero modes of this kind act only in a particular
part of the chain extending
from the left boundary to a certain position.
\\[10mm]
\noindent
{\bf Acknowledgements}\\[2mm]
I would like to thank V. Rittenberg
and I. Peschel for many valuable
discussions and A. Kr\"amer for reading the manuscript.
%
%
%
%
\appendix
\def\thesection {Appendix \Alph{section}}
\renewcommand{\theequation}{\Alph{section}.\arabic{equation}}
\section {The structure of fermionic zero modes}
\setcounter{equation}{0}
\noindent
The quantum group symmetry of the $XY$ chain
Hamiltonian (\ref{xy})
and its generalization (\ref{FourFermionHamiltonian}) relies
on the existence of fermionic zero modes for arbitrary
parameters. In this appendix we show that zero modes of this kind
possess a very simple and general structure. For this we derive the
zero mode condition (\ref{ZeroModeCondition}) on the level
of the diagonalization of the matrix $M$ in Eq. (\ref{EVP}).
\\
\indent
Since the matrix $M$ in Eq. (\ref{M}) possesses a tridiagonal
block structure it is near at hand to solve the system of
equations $M \vec{\psi}=0$ `line by line'. Introducing the
vector notation  $\vec{\psi}_{0,j}^\mu =
(\psi_{0,j}^{\mu,1},\ldots,\psi_{0,j}^{\mu,2n})$ and fixing
$\vec{\psi}_{0,1}^\mu$ on the leftmost site, the first line
of the matrix $M$ allows $\vec{\psi}_{0,2}^\mu$ to be computed by
\begin{equation}
\label{LeftBC}
\vec{\psi}_{0,2}^\mu \;=\; -A^{-1}B\,\vec{\psi}_{0,1}^\mu \,.
\end{equation}
\noindent
This equation can be understood as a left boundary condition.
In the same way the last line
\begin{equation}
\label{RightBC}
\vec{\psi}_{0,L-1}^\mu = {(A^T)}^{-1}C\,\vec{\psi}_{0,L}^\mu\,
\end{equation}
\noindent
acts as a right boundary condition. Between the boundaries the lines
$j=2,\,\ldots,L-1$
\begin{equation}
\label{Bulk}
\vec{\psi}_{0,j+1}^\mu \;=\; -A^{-1}(B+C)\,\vec{\psi}_{0,1}^\mu +
A^{-1}A^T\,\vec{\psi}_{0,j-1}^\mu
\end{equation}
describe the bulk behaviour of the zero mode. These equations
allow the zero mode eigenvector
to be constructed as follows. Once if $\vec{\psi}_{0,1}^\mu$
is fixed, Eq. (\ref{LeftBC}) yields $\vec{\psi}_{0,2}^\mu$.
Then by Eq.~(\ref{Bulk}) which can be understood as
a two-stage recurrency relation the vectors $\vec{\psi}_{0,3}^\mu,
\ldots,\vec{\psi}_{0,L}^\mu$
(and therewith the entire zero mode vector)
can be computed recursively. Finally the last two vectors
$\vec{\psi}_{0,L-1}^\mu$ and $\vec{\psi}_{0,L}^\mu$
have to satisfy the
right boundary condition Eq. (\ref{RightBC}). Normally
it is rather difficult to determine $\vec{\psi}_{0,1}^\mu$ so
that the right boundary condition~({\ref{RightBC}) is satisfied
since the parameters of the Hamiltonian $H(A,B,C)$
are involved in a highly non-linear way.
\\
\indent
The situation becomes much simpler
if we impose the additional condition
\begin{equation}
A^{T}+CA^{-1}B=0 \nonumber
\end{equation}
\noindent
on the matrices $A$, $B$ and $C$. The two-stage
recurrency relation (\ref{Bulk})
then reduces to a one-stage
relation $\vec{\psi}_{0,j}^\mu \;=\; -A^{-1}B\,
\vec{\psi}_{0,1}^\mu$ yielding solution in Eq. (\ref{PowerLaw}).
It is easy to see that these eigenvectors
satisfy the boundary conditions (\ref{LeftBC}) and
(\ref{RightBC}) {\it automatically}.

%
%
%
%
\section{Discrete Symmetry Transformations}
\setcounter{equation}{0}
Beside the quantum group invariance
the anisotropic $XY$ chain (\ref{xy}) possesses
a further important symmetry. Diagonalizing the
Hamiltonian (\ref{xy}) one observes that the exchange
of the parameters $\eta$ and $q$ does not modify the spectrum:
\begin{equation}
\label{SimH}
H^{XY}(\eta,q) \; \doteq \; H^{XY}(q,\eta)\;
\end{equation}
In Ref. \cite{Haye2} we derived the corresponding similarity
transformation
\begin{equation}
\label{udef}
H^{XY}(\eta,q) \;=\; U\,H^{XY}(q,\eta)\,U^{-1}
\end{equation}
\noindent
and showed that $U$ reduces in a special limit
to the Ising duality transformation. For this reason we denoted
the transformation (\ref{udef}) as
`generalized duality transformation'
(although the $XY$ chain is not self-dual in the usual sense).
In this section we show that a
similar symmetry exists in the
case of generalized chains of the form (\ref{GeneralHamiltonian})
obeying the zero mode condition (\ref{ZeroModeCondition}).
\\
\indent
We first notice that in the $XY$ chain case
the transformation~(\ref{SimH}) just inverts the deformation parameter
$\alpha_1 \leftrightarrow \alpha_1^{-1}$ while $\alpha_2$ is not changed
(in the same way it is possible to construct a similarity
transformation which inverts $\alpha_2$ and keeps $\alpha_1$ fixed).
Then looking at the fermionic excitation
energies (\ref{FourFermionExcitations})
of the generalized Hamiltonian (\ref{FourFermionHamiltonian})
with two fermionic degrees of freedom per site
we recognize that the inversion of any deformation parameter
$\alpha_\mu \leftrightarrow \alpha_\mu^{-1}\; (\mu=1,\ldots,4)$
does not alter the spectrum. We therefore expect
this observation to hold for arbitrary $n$,
i.e. using the notation of Eq. (\ref{SpecialChoice}) we assume that
for every $\mu=1,\ldots,2n$ we have
\begin{equation}
H(\alpha_1,\ldots,\alpha_\mu^{-1},\ldots,\alpha_{2n},\,\Omega)
\;\doteq\;
H(\alpha_1,\ldots,\alpha_\mu,\ldots,\alpha_{2n},\,\Omega)\,.
\end{equation}
\noindent
In analogy to the results of Ref. \cite{Haye2} it turns out that
the corresponding similarity transformation
depends exclusively on
the deformation parameter it is inverting:
\begin{equation}
H(\alpha_1,\ldots,\alpha_\mu^{-1},\ldots,\alpha_{2n},\,\Omega)
\;=\;
U(\alpha_\mu) \,
H(\alpha_1,\ldots,\alpha_\mu,\ldots,\alpha_{2n},\,\Omega)\,
U^{-1}(\alpha_\mu)\,.
\end{equation}
\noindent
Denoting
\begin{equation}
N_\mu\;=\;2^{L-1}\,\frac{1+\alpha_\mu^L}{(1+\alpha_\mu)^L}\;
\hspace{2cm}
\omega_\mu \;=\; \frac{\alpha_\mu^{1/2}-\alpha_\mu^{-1/2}}
{\alpha_\mu^{1/2}+\alpha_\mu^{-1/2}}
\end{equation}
\noindent
this transformation can be written in terms of a
`time-ordered' exponential
\begin{equation}
\label{trafo}
U(\alpha_\mu) \;=\; \frac{1}{\sqrt{N_\mu}}\,
\,T\,\exp(\omega_\mu G_\mu)\,,
\end{equation}
\noindent
where $G_\mu$ is a nonlocal generator
\begin{equation}
G_\mu \; = \sum_{1 \leq j_1<j_2 \leq L}
\tau^\mu_{j_1}\tau^\mu_{j_2}\,.
\end{equation}
\noindent
$T$ is an ordering operator defined by
\begin{equation}
T\,\tau_i^\mu\tau_j^\mu \;=\;
\left\{ \begin{array}{ll}
\hspace{2.3mm}\tau_i^\mu\tau_j^\mu\hspace{5mm}
& i<j \\
-\tau_j^\mu\tau_i^\mu & i>j \\
\hspace{5mm}0 & i=j
\end{array} \right\}\,.
\end{equation}
\noindent
Explicitely the transformation $U(\alpha_\mu)$ is given by
the polynomial
\begin{equation}
\label{OrthogonalDefinition}
U(\alpha_\mu)\;=\;\frac1{\sqrt{N_\mu}}\,\Biggl( \, {\bf{1}}\,+\,
\sum_{k=0}^{[L/2]}\,
\omega_\mu^k\sum_{1 \leq j_1<j_2<\ldots<j_{2k} \leq L}
\tau^\mu_{j_1}\tau^\mu_{j_2}\ldots\tau^\mu_{j_{2k}}\,\Biggr)
\end{equation}
where $[L/2]$ denotes the truncation of $L/2$ to an integer number. It is
an orthogonal transformation
\begin{equation}
U(\alpha_\mu)^T\;=\;U(\alpha_\mu)^{-1}
\end{equation}
\noindent
and its inverse is given by
\begin{equation}
U(\alpha_\mu)^{-1} \;=\; U(\alpha_\mu^{-1})\,.
\end{equation}
\noindent
Therefore $U(\alpha_\mu)$ reduces to the identity
if the deformation parameter in question is equal to one. Because
of $[G_\mu,G_\nu]=0$ the transformations $U(\alpha_\mu)$ commute
for different $\mu$ and can be combined freely. Notice that for
non-generic deformation parameters ($\alpha_\mu^L=\pm 1$)
the transformation $U(\alpha_\mu)$ does not exist since the
normalization $N_\mu$ diverges.
\\
\indent
It is well known in the theory of quantum groups that the inversion
of a deformation parameter corresponds to an algebra homomorphism. The
transformation (\ref{trafo}) shows that the same is true for the
whole physical system. It should be emphasized that
different physical situations
are related (e.g. disordered and frozen states).
If the deformation parameter in question is equal to one, the system
undergoes a massless phase transition. Two types of transitions are
possible. If the dispersion of the massless excitations is linear
in $k$, we have a criticial Ising transition, otherwise if the
dispersion is quadratic in $k$, we observe a Pokrovsky-Talapov
phase transition.
%
%
%
%

\vspace{15mm}
\noindent
{\bf Figure Captions:}

\noindent
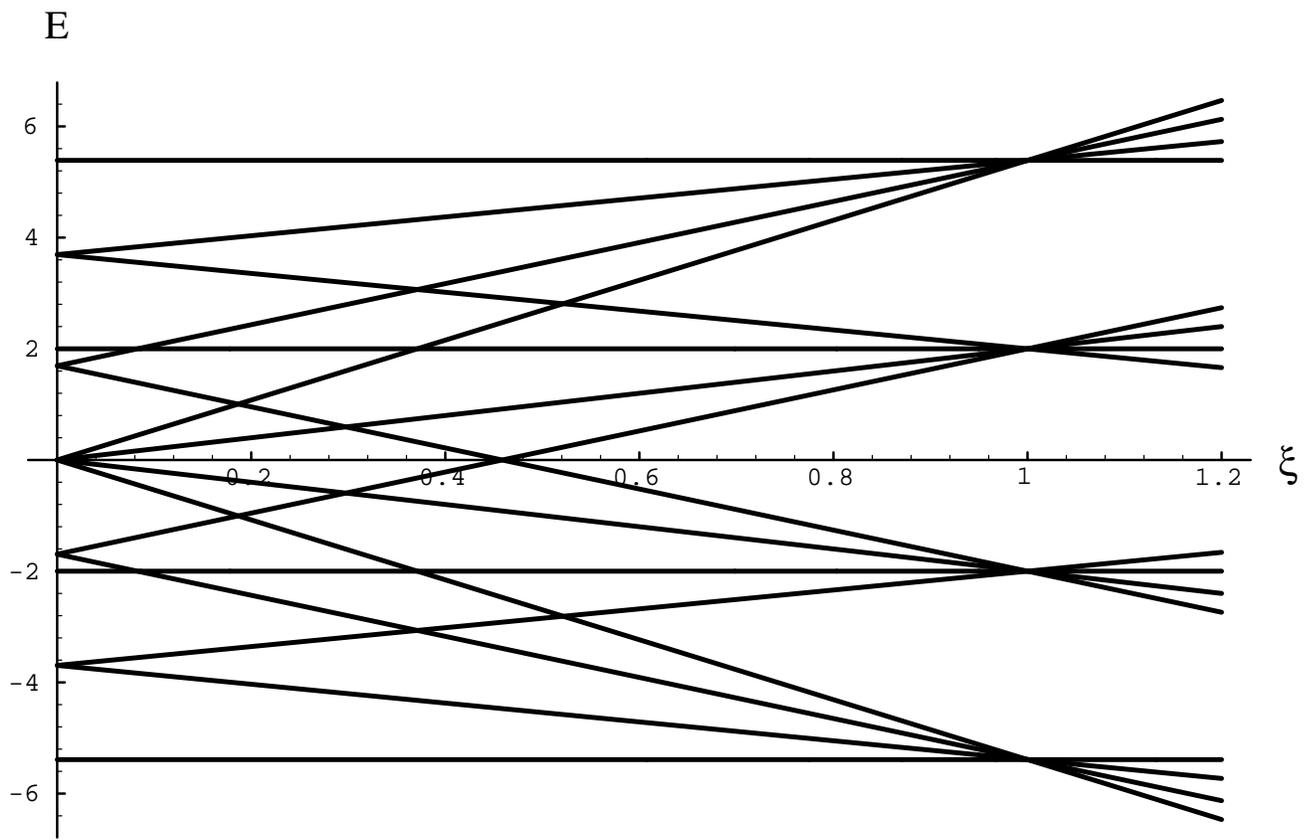
Fig. 1: Spectrum of $H(\frac23,\frac23,\frac23,\frac23,\xi)$ for 3 sites.

\noindent
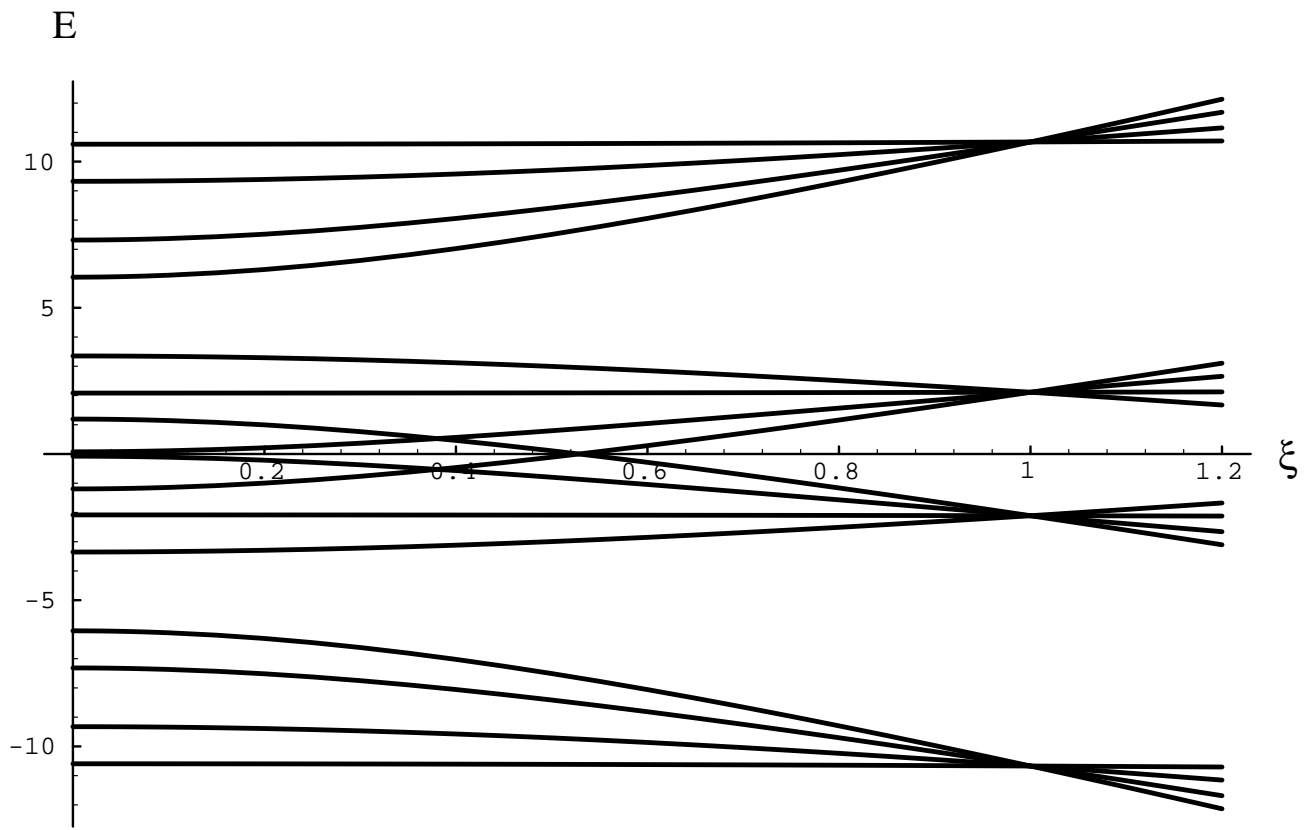
Fig. 2: Spectrum of $H(\frac27,\frac37,\frac57,\frac67,\xi)$ for 3 sites.

\end{document}